\documentclass[twocolumn,amsmath,amssymb,showpacs,aps,superscriptaddress]{revtex4-1}

\usepackage{graphicx}
\usepackage{bm}
\usepackage{amsmath}
\usepackage{amsfonts}
\usepackage{mathrsfs}

\begin{document}

\title{Probing polaron excitation spectra in organic semiconductors by photoinduced-absorption-detected two-dimensional coherent spectroscopy}

\author{Hao \surname{Li}}
\email{hli39@uh.edu}
\affiliation{Department of Chemistry, University of Houston, Houston, TX 77204, United States}

\author{Aur\'elie \surname{Gauthier-Houle}}
\affiliation{Department of Physics and Regroupement qu{\'e}b{\'e}cois sur les mat{\'e}riaux de pointe, Universit{\'e} de Montr{\'e}al, C.P. 6128, Succursale centre-ville, Montr{\'e}al  H3C~3J7, Canada}

\author{Pascal \surname{Gr\'egoire}}
\affiliation{Department of Physics and Regroupement qu{\'e}b{\'e}cois sur les mat{\'e}riaux de pointe, Universit{\'e} de Montr{\'e}al, C.P. 6128, Succursale centre-ville, Montr{\'e}al  H3C~3J7, Canada}

\author{Eleonora \surname{Vella}}
\affiliation{Department of Physics and Regroupement qu{\'e}b{\'e}cois sur les mat{\'e}riaux de pointe, Universit{\'e} de Montr{\'e}al, C.P. 6128, Succursale centre-ville, Montr{\'e}al  H3C~3J7, Canada}

\author{Carlos \surname{Silva-Acu\~na}}
\email{carlos.silva@umontreal.ca}
\affiliation{Department of Physics and Regroupement qu{\'e}b{\'e}cois sur les mat{\'e}riaux de pointe, Universit{\'e} de Montr{\'e}al, C.P. 6128, Succursale centre-ville, Montr{\'e}al  H3C~3J7, Canada}

\author{Eric R.\ Bittner}
\email{bittner@uh.edu}
\affiliation{Department of Chemistry, University of Houston, Houston, TX 77204, United States}

\date{\today}

\begin{abstract}
We report a theoretical description and experimental implementation of a novel two-dimensional coherent excitation spectroscopy based on quasi-steady-state photoinduced absorption measurement of a long-lived nonlinear population. We have studied a semiconductor-polymer:fullerene-derivative distributed heterostructure by measuring the 2D excitation spectrum by means of photoluminescence, photocurrent and photoinduced absorption from metastable polaronic products. We conclude that the photoinduced absorption probe is a viable and valuable probe in this family of 2D coherent spectroscopies.
\end{abstract}


\maketitle

\section{Introduction}
\label{sec:intro}

Multidimensional electronic spectroscopy (ES) developed in analogy to the counterpart of nuclear magnetic resonance has become a powerful tool to study the electronic dynamics of macromolecular systems~\cite{Branczyk:2014gf,Fuller:2015fk,Nardin:2016ul}. By probing nonlinear optical responses and spreading them along two time/frequency axes, such spectroscopic experiments reveal transient correlations between specific electronic transitions with femtosecond temporal resolution~\cite{mukamel1995principles,Mukamel:AnnuRevPC2000}. Usual implementations of 2D ES rely upon  optical wavevector matching to measure a third-order mesoscopic polarization response~\cite{Fuller:2015fk}.In this work, we focus upon the two-dimensional photoexcitation spectroscopy (2D PES), as a variant of 2D ES techniques, in which the detected nonlinear optical signals rely on the fourth-order excited-state population. According to the final population of interest, one can probe the excitation spectral signals in forms of photoluminescence (PL)~\cite{Marcus:2DFS:JCP2007,AHMarcus:PNAS2011,AHMarcus:2DFS-2012JPCB} or photocurrent (PC)~\cite{Nardin:2013uq,AHMarcus:NatCommun2014,Vella:2016zm}, for example. Indeed, this implementation of 2D coherent spectroscopy has the formidable advantage that it can exploit photocurrent detection as an extraordinarily sensitive population probe~\cite{Bakulin:2016lq}. In this contribution, we demonstrate an alternative approach that is complimentary to PL and PC detection schemes, based on quasi-steady-state photoinduced absorption (PIA). In this implementation, if the final population produced with the phase-modulated ultrashort pulses is long-lived on the timescale of the phase modulation period, it can be probed via an excited-state absorption by a continuous-wave probe beam. Here, we demonstrate the technique by comparing 2D coherent excitation spectra in a semiconductor-polymer:fullerene blend measured by PL, PC PIA.

Regarding the 2D ES implementations, the well-developed four-wave mixing (FWM) approach is a conventional technique in which non-collinear optical pulses are often employed~\cite{Fuller:2015fk}. Owing to advances in ultrafast optics, one can prepare precise sequences of pulses with well-defined pulse envelopes, wave-vector directions, time intervals, and relative phases. In FWM experiments, the first three laser pulses generate third-order polarization signals that obey wave-vector phase-matching conditions and then can be heterodyne-detected by a fourth pulse serving as a local oscillator~\cite{GRFleming:Accounts2009}. Measuring the polarization in the wave-vector-matched direction has the advantage of spatially ruling out undesired signals, such as the pure rephasing signal obtained in photon echo spectroscopy~\cite{DJonas:JCP2001}.

In 2D PES, a train of four collinear laser pulses subjected to precise phase-modulation is applied to the material system, with all the four pulses involved in the field-matter interaction. Therefore an excited-state population is created and the resulting fourth-order response can be obtained by measuring physical observable associated with excited-state population as an incoherent process. The phase-sensitive detection differentiates the signal into rephasing and non-rephasing ones which are analogous to photo echo and virtual echo counterparts in 2D ES, respectively, and it is possible to measure both responses with a multichannel lock-in amplifier. The collinear beam geometry used in this approach has several advantages over the wave-vector matched counterparts. For instance, the phase-matching condition $\Delta \bm k l \ll \pi$ requires larger interaction length $l$ in the FWM experiments, where $\Delta \bm k$ is the difference between wave-vector sum of the incoming laser fields and that of the polarization emission. Without such constraint, the spectroscopy using collinear pulses can be applied to samples of small size technically to the scale of single molecules \cite{Warren:JPCA1999,Tian:Science2003,Hulst:Nature2010}. In addition, due to the great simplification in the optomechanical setup, the phase-selective technique make it practically capable to probe nonlinear response of higher dimensions involving more field-matter interactions. 

\begin{figure}[t]
  \begin{center}
    \includegraphics[width=0.4\textwidth]{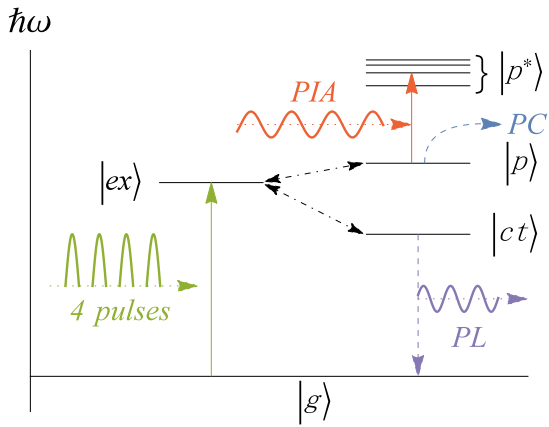}
  \end{center}
  \caption{Energy-level diagram of the four-level model describing photocurrent generation in polymer:fullerene systems. The exciton on the polymer ($|ex\rangle$) is produced by the collinear sequence of femtosecond pulses. It dissociates to produced polarons ($|p\rangle$), some of which can become photocarriers and contribute photocurrent. The lowest-energy ones are charge-transfer excitons ($|ct\rangle$), which have a radiative decay pathway yielding red-shifted, featureless emission spectra. Polaronic species can be thus probed via photoluminescence ($|ct\rangle$), photoinduced absorption (all charges), and photocurrent (photocarriers). 
  \label{fig:energy-diagram}}
\end{figure}

PIA spectroscopy is a quasi-steady state modulation technique (see~\citenum{Cardona:1969xy}) in which a continuous-wave (CW) modulated source generates long-lived photoexcitations, which are detected by changes in transmission of a probe lamp due to their excited-state absorption. This is a valuable spectroscopy to study slow kinetics of photoexitations such as polarons and triplet excitons in organic semiconductors, which are often outside of the windows that can be readily probed with ultrafast spectroscopies. In this work, we extend this approach to 2D PES by probing photoinduced fourth-order charge population in a semiconductor-pomymer:fullerene blend, generated by a sequence of four collinear, phase-modulated femtosecond pulses. A CW laser is tuned to the peak of the polaron photoinduced absorption band to probe the quasi-steady-state charge population. Fig.~\ref{fig:energy-diagram} depicts schematically the three different probes of charge population that can be measured in parallel to probe charges in these systems: 2D~PL, 2D~PC, and 2D~PIA.

\section{Theory of 2D PIA spectroscopy}
\label{sec:theory}


Usually a collinear sequence of four ultrashort pulses is used in the 2D PES to excite an electronic system. The relative phase of the pulses can be modulated by an acousto-optic Bragg cell, and the nonlinear signal, in principle, can be any physical observable associated with the excited-state population such as photoluminescence and photocurrent. 
The third-order material polarization is concerned in the conventional 2D ES because the system is driven only by the first three pulses and the fourth pulse is only used for detection of the weak polarization emission. In the 2D PES, the signal is associated with the excited-state population expanded to the fourth-order perturbation as all of the four pulses drive the field-matter interaction to generate the final excitation population.
Here we briefly recapitulate the theory on the 2D PES. The Hamiltonian of an electronic system interacting with optical fields can be written as $\hat{H}(t)=\hat H_0 + \hat H'(t)$, in which $\hat H_0$ is the time-independent material Hamiltonian. The perturbation Hamiltonian under the dipole approximation is given by $\hat H'(t)=-E({\bm r},t)\cdot {\hat \mu}$ with $E(\bm r,t)$ and $\hat \mu$ being the applied optical field at position $\bm r$ and the dipole operator, respectively. According to the number of radiation-matter interactions, one can expand the density operator, to satisfy the Liouville-von Neumann equation, as $\hat{\rho}_{\rm I}(t)=\hat{\rho}^{(0)}_{\rm I}(t)+\hat{\rho}^{(1)}_{\rm I}(t)+\hat{\rho}^{(2)}_{\rm I}(t)+...$ in the interaction picture hereafter denoted by the subscript I. An ordinary operator $\hat O$ can be transformed to the interaction representation by
\begin{align}
 \label{eqn:interaction_pic}
  \hat O_{\rm I}(t) \equiv \hat U_0^\dagger(t,t_0) \hat O \hat U_0(t,t_0),
\end{align}
with
\begin{align}
 \label{eqn:u0}
  \hat U_0(t,t_0) \equiv e^{-\frac{i}{\hbar}\hat H_0 (t-t_0)}
\end{align}
being the time evolution operator from $t_0$ to $t$ in case of the time-independent Hamiltonian $\hat H_0$.
The $n$th-order perturbative expansion of the density operator can be written as
\begin{align}
 \label{eqn:density}
 \nonumber
  \hat\rho^{(n)}_{\rm I}(t)= & \left(-\frac{i}{\hbar}\right)^n\int_{t_0}^{t} d\tau_n \int_{t_0}^{\tau_n} d\tau_{n-1} ... \int_{t_0}^{\tau_2} d\tau_1
  \\ & [\hat H'_{\rm I}(\tau_n),[\hat H'_{\rm I}(\tau_{n-1}),[...,[\hat H'_{\rm I}(\tau_1),\hat \rho_{\rm I}(t_0)]  ...]]],
\end{align}
where the perturbation in the interaction picture is given by $\hat H'_{\rm I}(t) \equiv \hat U_0^\dagger(t,t_0) \hat H'(t) \hat U_0(t,t_0) = - E(\bm r,t) \cdot \hat \mu_{\rm I}(t)$.
Utilizing the perturbative expansion of the density operator, the expectation value of an arbitary observable can be expanded accordingly as $\langle \hat O^{(n)}(t) \rangle \equiv {\rm Tr}[\hat O(t) \hat \rho^{(n)}(t)]={\rm Tr}[\hat O_{\rm I}(t) \hat \rho_{\rm I}^{(n)}(t)]$. Consequently, the third-order polarization detected in the conventional 2D ES is given by $P^{(3)}={\rm Tr}[\hat \mu_{\rm I}(t) \hat \rho_{\rm I}^{(3)}(t)]$ and the corresponding response function $S^{(3)}(t_3,t_2,t_1)$ can be thereafter determined. 

Introducing the projection operator $\hat A=\sum_i |i\rangle\langle i|$ over all excited states $|i\rangle$ of interest, the total population of these states is given by the corresponding diagonal elements of the density operator
\begin{align}
 \label{eqn:population}
  \sum_i \rho^{(n)}_{ii}(t)=\langle \hat A^{(n)}_{\rm I}(t)\rangle={\rm Tr}\left[\hat A_{\rm I}(t)\hat \rho^{(n)}_{\rm I}(t)\right].
\end{align}
Analogous to the definition of the nonlinear polarization response functions, we can express the $n$-th order population response function with respect to the number of perturbations using the projector $\hat A$ instead of $\hat \mu$ as
\begin{widetext}
\begin{align}
 \label{eqn:pop-response}
   S^{(n)}(t_n,t_{n-1},...,t_1)=\left(\frac{i}{\hbar}\right)^n {\rm Tr} \left\{ \hat A_{\rm I}(t_n+t_{n-1}+...+t_1)\left[\hat \mu_{\rm I}(t_{n-1}+...+t_1),\left[...\left[\hat \mu_{\rm I}(t_1),[\hat \mu_{\rm I}(0),\hat \rho(-\infty)] \right]...\right]\right] \right\},
\end{align}
\end{widetext}
where $t_1, t_2, ..., t_n$ stand for the time intervals along the time-ordered radiation-matter interactions. Because state population does not change sign when the direction of all field vectors is reversed, only the response functions of even order are nontrivial, which is contrary to the polarization-probed spectroscopy where signals of even order disappear, e.g., the 2D ES measures the third-order polarization whereas the 2D PES probes the fourth-order population. Based on Eq.~(\ref{eqn:pop-response}), correlation functions can be explicitly written and the corresponding Liouville pathways can be determined according to the excitation model.

In the case of single excitation manifold, the fourth-order perturbation in Eq.~(\ref{eqn:pop-response}) results in opposite sign for both ground-state bleaching (GSB) and stimulated emission (SE) pathways comparing to the third-order expansion of 2D ES. In models with multiple excitation manifolds, significant difference in spectra between the 2D~ES and the 2D~PES is originated from how to take into account the Liouville pathways of excited-state absorption (ESA), whose contribution to the spectra is regulated by both the commutator expansion (determining the sign) and the population ending in higher excited-state manifold (altering the magnitude).



We have previously investigated the 2D~PC spectra in polymeric solar cells and proposed a four-level model to describe the process of free charge generation \cite{Vella:2016zm}. Intuitively, the 2D~PC spectroscopy is a straightforward method to explore the coupling between excitonic and free charge states involved in the charge separation process in molecular photovoltaics. As depicted in Figure~\ref{fig:energy-diagram}, the electron system is first excited from the ground state $|g\rangle$ to the exciton manifold $|ex\rangle$. Under the quantum coherence induced by the environment, the exciton state is coupled to the bound charge transfer (CT) state $|ct\rangle$ pinned at the heterojunction interface, and to the polaron manifold $|p\rangle$ ultimately responsible for the photocurrent generation.

In previous work \cite{Vella:2016zm}, we generally assumed that the strength of the photocurrent signal was proportional to the polaron population without distinguishing the internal structure of the polaron manifold. Due to the large scale of charge separation, the lifetime of a polaron is much longer than that of an exciton. Therefore, one can conveniently apply the CW radiation, that matches the resonance condition of $|p\rangle \rightarrow |p^*\rangle$ with $|p^*\rangle$ standing for the excite polaron manifold, to measure the photoinduced change in the absorption spectrum. 
Here we combine the CW and the time-resolved measurements, and propose a complimentary technique for the 2D~PES where the observable is the PIA of long-lived species. Again, the four collinear ultrashort pulses allow to probe the electronic coherence, but the fourth-order population is now detected with the change in transmission of a CW laser caused by PIA of polaron and CT states. Furthermore, the nonlinear response attributed to the CT state population can be measured by the 2D~PL spectroscopy.


In the presence of CW optical source, the Hamiltonian accounting for the CW interaction is given by $\hat H_{\rm cw}(t)=-E_{\rm cw}(\bm r,t)\cdot \hat \mu$. We consider the presence of the CW laser together with the material Hamiltonian while the pulse radiation-matter interactions are still being treated as perturbations which allow for the identical solution of the Liouville-von Neumann equation as in Eq.~(\ref{eqn:density}). The problem therefore turns out to be the 2D~PES under a material Hamiltonian with time-dependent off-diagonal elements attributed to $\hat H_{\rm cw}(t)$. Within the interaction picture, the expansion of the density operator is identical to Eq.~(\ref{eqn:density}). 
However, under such time-dependent Hamiltonian $\hat H_0 + \hat H_{\rm cw}(t)$, $\hat U_0(t,t_0)$ defined in Eq.~(\ref{eqn:u0}) is inadequate to describe the system evolution. So in the consideration of the CW radiation, $\hat U_{0,\rm cw}(t,t_0)$, the time evolution operator without interacting with optical pulses obeys
\begin{align}
 \label{eqn:u0cw-eq}
  \frac{\partial}{\partial t} \hat U_{0,\rm cw}(t,t_0)=-\frac{i}{\hbar} [\hat H_0 +\hat H_{\rm cw}(t)] \hat U_{0,\rm cw}(t,t_0)
\end{align}
and can be rewritten as
\begin{align}
 \label{eqn:u0cw}
  \hat U_{0,{\rm cw}}(t,t_0) \equiv \hat U_0(t,t_0) \hat U_{\rm cw}(t,t_0),
\end{align}
in which $\hat U_{\rm cw}(t,t_0)$ is the evolution operator with respect to $\hat H_{\rm cw}(t)$ and satisfies
\begin{align}
 \label{eqn:ucw}
  \frac{\partial}{\partial t} \hat U_{\rm cw}(t,t_0)=-\frac{i}{\hbar} \hat H_{\rm cw,I}(t) \hat U_{\rm cw}(t,t_0)
\end{align}
with $\hat H_{\rm cw,I}(t) \equiv \hat U_0^\dagger (t,t_0) \hat H_{\rm cw}(t) \hat U_0 (t,t_0)$ being the CW Hamiltonian in the interaction picture with respect to $H_0$. In principle, the newly defined time evolution operators $\hat U_{0,\rm cw}(t,t_0)$ and $\hat U_{\rm cw}(t,t_0)$ can be expanded in a {\it Dyson series} but cannot be approximately truncated at lower orders.

The density operator can still be written according to the number of perturbations (optical pulses) as in Eq.~(\ref{eqn:density}) and the $n$-th order population response function remains the same as Eq.~(\ref{eqn:pop-response}). The difference is that, in the presence of CW radiation, an operator in the interaction representation is now the unitary transformation of $\hat U_{0,\rm cw}(t,t_0)$ given by
\begin{align}
 \label{eqn:interaction_pic_cw}
  \hat O_{\rm I}(t) = \hat U_{0,\rm cw}^\dagger(t,t_0) \hat O \hat U_{0,\rm cw}(t,t_0),
\end{align}
instead of using $\hat U_0(t,t_0)$ in Eq.(\ref{eqn:interaction_pic}).

It is worth mentioning that we apply unitary transformations twice in the aforementioned strategy. First the CW Hamiltonian is transformed by $\hat U_0(t,t_0)$ so that one can determine the time evolution operator $\hat U_{0,\rm cw}(t,t_0)$. Then the propagation of the dipole operator and the state projector in time can be found from Eq.~(\ref{eqn:interaction_pic_cw}). However, the perturbation expansion is only applied to the optical pulses but not the CW radiation. This assures such expansion of density operator can hold for long times even truncated at low orders.

We assume that the CW laser acts before the first optical pulse, set at $t=0$, and continues throughout the experiment. Because there is no population prior to $t=0$, the CW laser has no effect on the system so that the system Hamiltonian can be represented by $\hat H_0$ and the system is in a stationary state $\rho(-\infty)=\rho(0)$.

The fourth-order response function of the 2D~PES can be written as
\begin{align}
 \label{eqn:4th-reponse}
  S^{(4)}(t_4,...,t_1)=\left(\frac{i}{\hbar}\right)^4 \sum_{\alpha=1}^{8} \left[Q_{\alpha}(t_4,...,t_1)+Q^*_{\alpha}(t_4,...,t_1)\right]
\end{align}
in terms of correlation functions. Based on the model of single excitation manifold (hereafter denoted by subscript $a$) described in Figure~\ref{fig:energy-diagram}, the four possible Liouville pathways depicted in \cite{Vella:2016zm} are corresponding to two GSBs and two SEs and the correlation functions are
\begin{widetext}
\begin{align}
 \label{eqn:rp}
  Q_{3a}(t_4,t_3,t_2,t_1)=-{\rm Tr}\left[\hat A_{\rm I}(t_4+t_3+t_2+t_1) \hat \mu_{\rm I}(t_1) \rho(-\infty) \hat \mu_{\rm I}(0) \hat \mu_{\rm I}(t_2+t_1) \hat \mu_{\rm I} (t_3+t_2+t_1)\right],\\
  Q_{4a}(t_4,t_3,t_2,t_1)=-{\rm Tr}\left[\hat A_{\rm I}(t_4+t_3+t_2+t_1) \hat \mu_{\rm I}(t_2+t_1) \rho(-\infty) \hat \mu_{\rm I}(0) \hat \mu_{\rm I}(t_1) \hat \mu_{\rm I} (t_3+t_2+t_1)\right],\\
  Q_{2a}(t_4,t_3,t_2,t_1)=-{\rm Tr}\left[\hat A_{\rm I}(t_4+t_3+t_2+t_1) \hat \mu_{\rm I}(0) \rho(-\infty) \hat \mu_{\rm I}(t_1) \hat \mu_{\rm I}(t_2+t_1) \hat \mu_{\rm I} (t_3+t_2+t_1)\right],\\
  Q_{5a}^*(t_4,t_3,t_2,t_1)=-{\rm Tr}\left[\hat A_{\rm I}(t_4+t_3+t_2+t_1) \hat \mu_{\rm I}(t_2+t_1) \hat \mu_{\rm I}(t_1) \hat \mu_{\rm I}(0) \rho(-\infty) \hat \mu_{\rm I} (t_3+t_2+t_1)\right].
\end{align}
\end{widetext}
Here $Q_{4a}$ and $Q^*_{5a}$ are attributed to the GSB pathways while $Q_{2a}$ and $Q_{3a}$ are for SE ones. The notation and numbering of correlation functions follow from the reference \cite{AHMarcus:2DFS-2012JPCB}. According to the phase combination, the correlation functions can be grouped as rephasing signal $Q_{3a}$ and $Q_{4a}$ with phase $\phi_{21}-\phi_{43}$, and non-rephasing part $Q_{2a}$ and $Q^*_{5a}$ with phase $-(\phi_{21}+\phi_{43})$. The time evolved operators $\hat A_{\rm I}(t)$ and $\hat \mu_{\rm I}(t)$ can be determined from Eqs.~(\ref{eqn:u0cw-eq}-\ref{eqn:interaction_pic_cw}) by setting $t_0=0$.
It is worth mentioning that in case of a multiple excitation manifold (denoted by subscript $b$), the ESA pathways associated with the correlation functions $Q^*_{2b}$ (rephasing) and $Q^*_{3b}$ (non-rephasing) result in different spectra in the measurement of PL, PC and PIA because 
each uses different final state projectors.
 The details of these differences will be discussed elsewhere.
 Therefore, by selecting proper detection technique, 2D 
photoexcitation spectroscopy can be used to 
investigate excitation dynamics involving a variety of different states of interest.

\section{Experimental methods}

\begin{figure}[t]
\begin{center}
\includegraphics[width=\columnwidth]{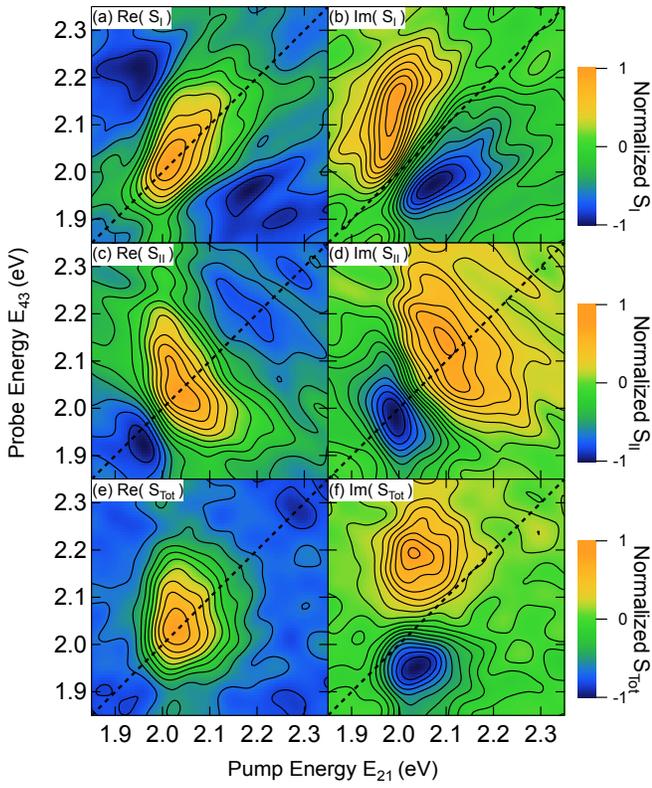}
\end{center}
\caption{2D PL measurements on a PCDTBT:PCBM film at ambient temperature, measuring the integrated PL of the charge-transfer exciton, at a 200 fs population time. Shown are the real (left column) and imaginary (right column) parts of the rephasing $S_I$ (a, b), non-rephasing $S_{II}$ (c,d), and total $S_{Tot}$ correlation function (e,f).
\label{fig:2DPL}}
\end{figure}
\begin{figure}[t]
\begin{center}
\includegraphics[width=\columnwidth]{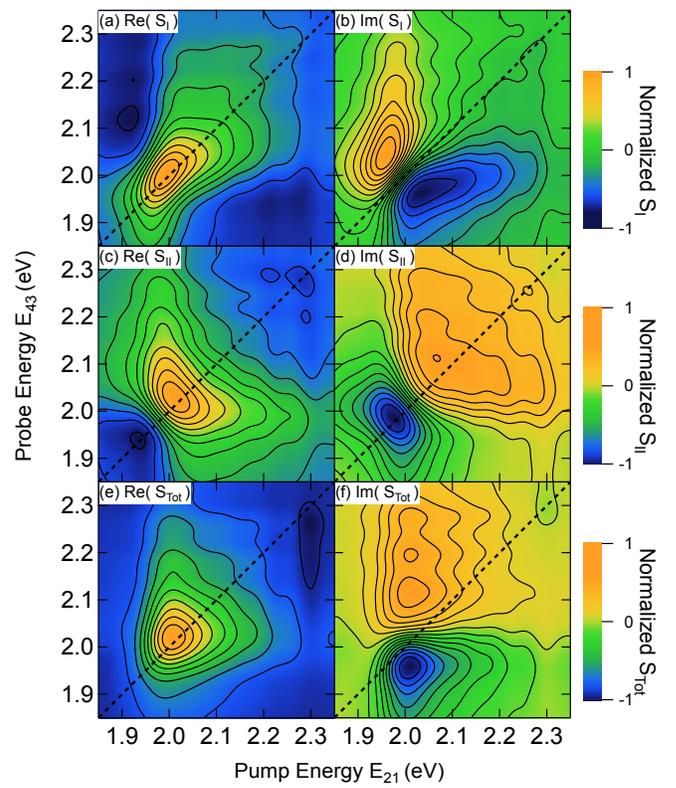}
\end{center}
\caption{2D PC measurements on a PCDTBT:PCBM photovoltaic diode (see ref.~citenum{Vella:2016zm}) at ambient temperature. The panels are analogous to those in Fig.~\ref{fig:2DPL}. The population time is 50 fs.
\label{fig:2DPC}}
\end{figure}
\begin{figure}[th]
\begin{center}
\includegraphics[width=\columnwidth]{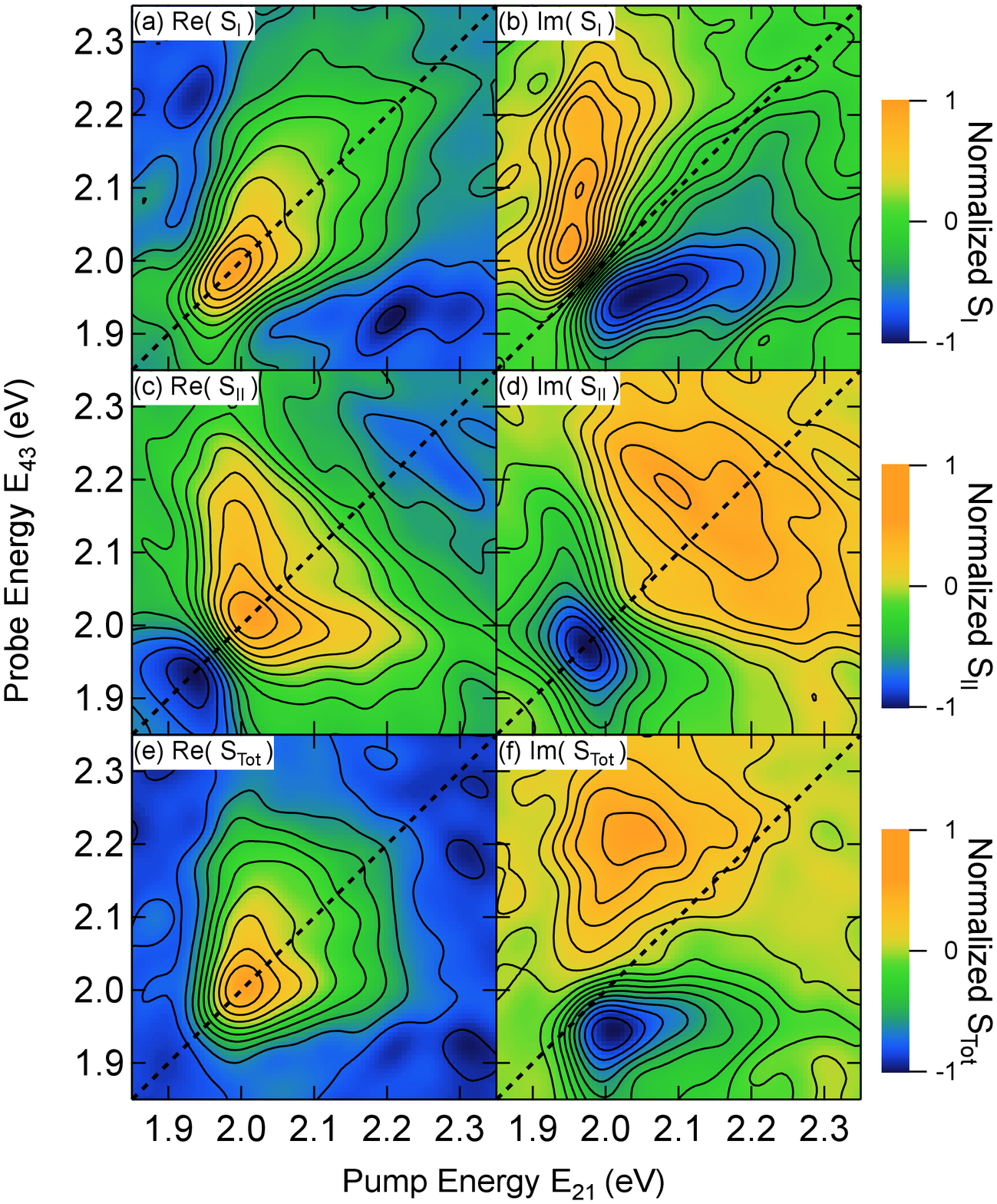}
\end{center}
\caption{2D PIA measurements on a PCDTBT:PCBM film at ambient temperature, with a population time of 100 fs. The panels are analogous to those in Fig.~\ref{fig:2DPL}. The CW laser probe energy was 1.61 eV.
\label{fig:2DPIA}}
\end{figure}

Our 2D~PES set-up is based on four collinear femtosecond pulses and acousto-optic phase modulation~\cite{Marcus:2DFS:JCP2007}. The four collinear pulses are generated using an apparatus described elsewhere~\cite{Vella:2016zm}. The output pulse train of a regenerative amplifier (Light Conversion Pharos, $\lambda_0$=1030$\,$nm, pulse duration $\sim 200$\,fs) operated at a repetition rate of 600\,kHz, pumps a Non-collinear Optical Parametric Amplifier (NOPA) delivering large bandwidth pulses ($\sim 80$ -- 100\,nm) centred at around 600\,nm. The pulses are compressed by an adaptive $4f$ pulse shaper (BioPhotonics Solutions FemtoJock-P) in order to be $\sim 10$-fs long at the sample position. By mean of a 50:50 beam-splitter (BS) the pulse train is then split to feed two twin Mach-Zehnder interferometers nested in an outer Mach-Zehnder interferometer. Three computer controlled delay stages allow independent control of the three interpulse delays $t_{21}$, $t_{43}$ and $t_{32}$. These time delays correspond to $t_1$, $t_3$, and $t_2$, respectively, in the preceeding section.  Each of the inner Mach-Zehnder interferometers produces a pair of phase-locked pulses. In each arm of the two twin interferometers, an acousto-optic Bragg cell modulates each pulse at an unique frequency ($\Omega_i$, $i=1$, 2, 3 and 4) close to 200$\,$MHz. The exact $\Omega_i$ values are chosen so that the difference between the modulation frequencies of the two pulses within a pair ($\Omega_{21}$ and $\Omega_{43}$) is of the order of $\sim 10$\,kHz.

For a detailed description of the phase modulation scheme adopted in the experiment, we refer the readers to Ref.~\citenum{Marcus:2DFS:JCP2007} (see also Ref.~\citenum{Nardin:2013uq} for a related approach), here we will briefly outline its basic principle. The time interval between consecutive laser pulses ($T_{\textrm{rep}}$) is set by the laser repetition rate, 600$\,$kHz. Due to the acousto-optic modulators, each of the pulses generated by the two twin Mach-Zehnder interferometers has a frequency shift imparted, equal to the respective frequency $\Omega_i$. Although these frequency changes are negligeable as compared to the optical frequency, 
they introduce a shift in the temporal phase of each pulse that oscillate at the corresponding frequency $\Omega_i$. 
At each laser shot, the experiment is repeated with a sequence of pulses phase-shifted with respect to the previous shot, thus creating two collinear trains of phase-modulated pulse pairs. Those interfering excitation pulses produce a population signal oscillating at $\Omega_{21}$ and $\Omega_{43}$ in the kHz range. It can be readily shown that~\cite{Marcus:2DFS:JCP2007}, within this phase modulation scheme, the non-linear signals of interest, which in analogy to four wave mixing experiments are referred to as \textit{rephasing} and \textit{non-rephasing} signals, oscillate at the frequencies $\Omega_{43}-\Omega_{21}$ and $\Omega_{43}+\Omega_{21}$ respectively. These frequency components are extracted simultaneously from the overall  signal by dual lock-in detection. The phase-sensitive detection scheme also offers a dinstinct advantage in the self-stabilization of the experiment. Optical replicas of the two pulse-pairs are generated at the exit beam splitters of the two inner Mach-Zehnder interferometers and used to generate the reference signals for the dual lock-in demodulation. The two sets of pulse replicas are sent to two monochromators, which spectrally narrow them (thus temporally elongating them), and are then detected by two avalanche photodiodes. The temporal elongation of the pulses provided by the monochromators produces reference signals for time delays $t_{21}$ and $t_{43}$ up to $\sim10$\,ps. It is important to note that, when scanning $t_{21}$, the phase of the references built in this way does not evolve at an optical frequency, but at a reduced frequency given by the difference between the frequency of the signal and that set by the monochromators. This frequency downshift results in an improvement of the signal-to-noise ratio, which is inversely proportional to the frequency downshift itself, and which virtually removes the impact of the mechanical fluctuations occurring in the setup on the signals of interest~\cite{Marcus:2DFS:JCP2007}. In order to obtain the reference signals at the frequencies of the rephasing and non-rephasing signals, $\Omega_{43}-\Omega_{21}$ and $\Omega_{43}+\Omega_{21}$, one of the two photodiode outputs (typically the one at higher frequency) undergoes amplitude modulation (AM) by the output of the other photodiode. The AM signal so obtained carries the two sideband frequencies of interest ($\Omega_{43}-\Omega_{21}$ and $\Omega_{43}+\Omega_{21}$) is then used for the lock-in demodulation.

The 2D maps are built acquiring the demodulated signals at fixed $t_{32}$ and by scanning $t_{21}$ and $t_{43}$; $t_{21}$ and $t_{43}$ are coherence times, and $t_{32}$ is the population waiting time. Specifically, for a given $t_{43}$, data are sequentially recorded at different $t_{21}$ in the interval of interest, typically extending to a few hundred femtoseconds, $t_{43}$ is then stepped, and the $t_{21}$ scan repeated until the full 2D time response is recorded. Each of such scans simultaneously produces four maps: the in-phase and the in-quadrature ones for the rephasing and non-rephasing frequencies. The maps so obtained in the time domain are, finally, converted to the energy domain by Fourier-transforming the time variables $t_{21}$ and $t_{43}$ and recorded as a function of the population waiting time, $t_{32}$. 

We have carried out three different implementations of 2D~PES on a 50\%wt film of the semiconductor polymer poly(N-90-heptadecanyl-2,7-carbazole-alt-5,5-(40,70-di-2-thienyl-20,10,30-benzothiadiazole)) as electron donor and of the fullerene derivative [6,6]-phenyl-C$_{60}$ butyric acid methyl ester as electron acceptor (PCDTBT:PC60BM). This is a workhorse semiconductor polymer heterostructure that delivers power conversion efficiencies in optimised solar cells greater than 6\%~\cite{Park:2009ys}. To perform 2D~PL measurements, we detect the charge-transfer exciton emission peaked at 1.61\,eV~\cite{Provencher:2012zl} by filtering the emission with a 700-nm long-pass filter. The detected signal is then sent as input to the lockin amplifier (Zurich Instruments HF2LI amplifier equipped with MF and MOD modules) and demodulated at the combination frequencies $\Omega_{43}-\Omega_{21}$ and $\Omega_{43}+\Omega_{21}$. For 2D~PIA, we use the same experimental setup as 2D~PL, except that we probe the photoinduced absorption at 1.61\,eV, characteristic of polarons on PCDTBT~\cite{Provencher:2012zl}, using a CW Ti:sapphire laser (Coherent Model 890), and detecting changes in transmission at the reference frequencies $\Omega_{43}-\Omega_{21}$ and $\Omega_{43}+\Omega_{21}$. The 2D~PC measurements were carried out using the same femtosecond excitation pulse sequence as above on a device structure described elsewhere~\cite{Vella:2016zm}. The photocurrent was amplified and converted to voltage (Zurich Instruments HF2TA Current amplifier), and was demodulated at the same reference combination frequencies. All measurements on films were carried out under vacuum of  $<10^{-3}$\,mbar and at ambient temperature.

\section{Discussion}

In this paper we have proposed and demonstrated a novel implementation of two-dimensional coherence excitation spectroscopy, in which a long-lived fourth-order population produced by a phase-modulated sequence of four femtosecond laser pulses is probed by its quasi-steady-state photoinduced absorption (PIA). We have compared 2D excitation spectra measured by photoluminescence of charge-transfer excitons, photocurrent, and polaron photoinduced absorption in a polymer:fullerene-derivative distributed heterostructure, and have found that the PIA detection scheme is a viable and potent probe of two-dimensional coherent excitation spectra of polarons, and adds to the arsenal of this class of techniques.

We have measured 2D~PL, 2D~PC, and 2D~PIA spectra on PCDTBT:PCBM, and these spectra are displayed, respectively, in Figs.~\ref{fig:2DPL}, \ref{fig:2DPC}, and \ref{fig:2DPIA}. In each case, the rephasing (a,b) and non-rephasing (c,d) spectra reveal a common 2D-excitation inhomogeneously broadened spectrum, indicating that charge-transfer excitons and photocarriers are produced from excitons on the polymer donor. The 2D~PIA spectrum probes \emph{all} polaron population that is characterised by the broad, featureless photoinduced absorption centred at 1.6\,eV~\cite{Provencher:2012zl}. The latter spectra are remarkably similar in lineshape to 2D~PL and 2D~PC spectra, as expected given that both photocarriers and charge-transfer excitons are polaronic in nature and carry the corresponding sub-gap optical signatures.

In previous publications, we have found by means of femtosecond stimulated Raman spectroscopy, that polaron generation in PCDTBT:PCBM occurs on timescales $\lesssim 100$\,fs~\cite{Provencher:2014fv}. We have proposed that this is dominated by a mechanism by which fluctuations in the coupling between the electronic system and its bath induce resonant tunnelling between various asymptotic states of the system, and that such noise-induced coherence could be a mechanism for populating delocalised polarons that are photocurrent-producing states~\cite{Bittner:NatCom2014}. Nevertheless, given the energy scale on which such off-diagonal couplings fluctuate, one might expect decoherence times to be ultrafast as well, which we conclude by modelling 2D~PC measurement using lower laser bandwidth in a previous publication~\cite{Vella:2016zm}. This would be true then for resonant tunneling between exciton and \emph{all} accessible polaron states, and so the spectral linewidth from 2D~PL, 2D~PC, and 2D~PIA measurements would then be expected to be similar. Comparison of Figs.~\ref{fig:2DPL}, \ref{fig:2DPC}, and \ref{fig:2DPIA} are consistent with this expectation.


It is important to point out that each of these spectroscopies (2D-PL, 2D-PC, and 2D-PIA) 
are based upon  4-th order population response functions that differ {\em only} in the final 
projection of the population.  In the 2D-PIA the projection is 
onto the manifold of excited polarons $|p^*\rangle$, in the 2D-PC, the projection is onto 
a set of out-going, photo-current producing states, and finally in 2D-PL, the projection is 
the radiation from the $|ct\rangle$ state as diagrammed in Fig.~\ref{fig:energy-diagram}.  Hence, all 
of these are complementary methods that bring additional insight into the fate of long-lived 
photo-excitations. 
It is also important to note that these methods are not solely restricted to photovoltaic systems and
could provide crucial insight into the excited state dynamics with of a wide range of systems
with long-lived photo-excitations including  those undergoing
triplet recombination and singlet exciton fission.

\begin{acknowledgements}
The work at the University of Houston was funded in part by the
National Science Foundation (CHE-1362006)
and the Robert A. Welch Foundation (E-1337). The work in Montreal was funded by the Natural Science and Engineering Research Council of Canada (NSERC), the Fonds de recherche du Qu\'ebec -- Nature et technologies (FRQ-NT), and CS's University Research Chair. PG acknowledges a Doctoral Postgraduate Fellowship from NSERC. We gratefully acknowledge Sachetan Tulhadar and Jenny Nelson for the device reported in Ref.~\citenum{Vella:2016zm}.

\end{acknowledgements}

\bibliography{bibliography}

\end{document}